\documentstyle[12pt,epsf]{article}
\def\largelinestretch{\renewcommand{\baselinestretch}{1.3}}
\largelinestretch\small\normalsize

\title{
\vspace*{-10mm}
\hfill
\parbox{4cm}{\large JINR E2-97-139\\
                    hep-ph/9704354}\\
\vspace*{10mm}
\bf Results on $K\to 2\pi$ decays at $O(p^6)$\\
    and $\varepsilon^{'}/\varepsilon$ from an effective\\
    chiral lagrangian approach}
 \author{
A.A.Bel'kov${}^1$\thanks{E-mail: {\tt belkov@cv.jinr.dubna.su}}~,
G.Bohm${}^2$\thanks{E-mail: {\tt bohm@ifh.de}}~,
A.V.Lanyov${}^{1,2}$\thanks{E-mail: {\tt lanyov@ifh.de}}~,
A.A.Moshkin${}^1$\thanks{E-mail: {\tt andrem@cv.jinr.dubna.su}}
\\[1ex]
\small
${}^1$
        Particle Physics Laboratory, Joint Institute for Nuclear
        Research,
\hfill\\[-0.2ex]
\small
        141980 Dubna, Moscow region, Russia
\hfill\\[0.2ex]
\small
${}^2$
DESY -- Zeuthen, Platanenallee 6, D-15735 Zeuthen, Germany
\hfill\\
}
\date{}

\begin{document}
%\largelinestretch\normalsize
\thispagestyle{empty}
\begin{titlepage}
\thispagestyle{empty}
\maketitle
\begin{abstract}
%\normalsize
    We have combined a new systematic calculation of mesonic matrix
elements at $O(p^6)$ from an effective chiral lagrangian approach using
Wilson coefficients taken from \cite{buras3}, derived in the framework
of perturbative QCD, and restricted partly by experimental data.
    We derive complete expressions for $K\to 2\pi$ amplitudes and give
new estimates for $\varepsilon^{'}/\varepsilon$.
\end{abstract}

\end{titlepage}

%============================================================================
%\begin{center}
%\large {\bf 1. Introduction}
%\end{center}
\section{Introduction}
%============================================================================
    The starting point for most advanced calculations of nonleptonic
kaon decays is an effective weak lagrangian of the form
\cite{vzsh,gilman-wise}
\begin{equation}
{\cal L}^{q}_{w}\big(|\Delta S|=1\big) =
\sqrt2\,G_F\,V_{ud}V^{*}_{us}\sum_{i} C_i \, {\cal O}_i
\label{weak-lagr}
\end{equation}
which can be derived with the help of the Wilson operator product
expansion from elementary quark processes, with additional gluon
exchanges.
   In the framework of perturbative QCD the coefficients $C_i$ are
to be understood as scale- and normalization scheme dependent
functions.
   There exist extensive next-to-leading order (NLO) calculations
\cite{buras1,ciuchini} in the context of kaon decays, among others.
   These calculations are based on the possibility of factorization of
short- and long-range contributions, i.e. into Wilson coefficient
functions $C_i$ and mesonic matrix elements of four-quark operators
${\cal O}_i$, respectively.
   The latter, however, can presently be obtained only by using
nonperturbative, i.e. model-dependent, methods.
   But still it seems not possible so far, to fulfil the
obvious requirement of scale- and renormalization scheme invariance of
the resulting amplitudes in a satisfactory way.

   Usually, the results of calculations are displayed with the help of
$B$-factors in the form
$$
T_{K\to 2\pi} = \sqrt2\,G_F\,V_{ud}V^{*}_{us} \sum_{i}
                 \Big[ C_i(\mu)B_i(\mu) \Big]
                 <\pi\pi |{\cal O}_i|K>_{vac.sat.}\,,
$$
where the mesonic matrix elements of four-quark operators are
approximated by their vacuum saturation values which are, of course,
$\mu$-independent.
   In principle, $B_i(\mu)$ should be estimated by some higher-order
calculations in the long-range regime, for instance, in
$1/N_c$-expansion \cite{buras2} in the form $1+O(1/N_c)$, or from the
lattice approach.
   The preliminary stage of these calculations is best characterized
by the problem to explain the well-known $\Delta I = 1/2$ rule
quantitatively.
   Of course, the lack of such calculations for long-range effects
severely restricts the predictive power of (\ref{weak-lagr}), leaving
only the possibility of some semi-phenomenological treatment
\cite{buras1,buras3,buras4}, with correspondingly large theoretical
uncertainties.
   As a matter of fact, some of the combinations
$\big[C_i(\mu)B_i(\mu)\big]$ have to be fixed by experimental data,
while others are restricted by additional theoretical arguments.

   A logical improvement of this approach, pursued in
\cite{bos-weak1,bos-weak2}, is to take the lagrangian (\ref{weak-lagr}) as
defining the structure of the $|\Delta S|=1$ weak interaction and to
submit it to a global confrontation with data by sandwiching it
between appropriate initial and final states (for instance, charged and
neutral kaons, respectively $2\pi$, $3\pi$, $n\pi\gamma$,... states).
   Of course, there would be hardly any gain in predictive power, if
we stuck to the vacuum saturation approximation because there may
be large unknown factors involved, when switching between different
initial/final states of the same operator.
   It is evident that everything could be absorbed into the $B_i$-factors
which thereby would not only be scheme- and scale-dependent, but would
also change with the initial/final states (becoming even unknown
(complex) functions of dynamic variables for final states).
   On the other hand, these latter changes should be avoided or, at
least, diminished when using a consistent higher-order calculation of
the mesonic matrix elements in the framework of chiral theory which
today is, certainly, the most developed approach to long-range hadronic
phenomena in general.
   These higher-order calculations after the inclusion of meson loops
have their own scale dependent renormalization ambiguities.
   In the ideal case these meson loops should cancel the scale dependence
of the Wilson coefficients from the perturbative QCD calculation
(resulting in $B$-factors equal to 1).
   As a complete match is lacking so far, the best one can do is
to fix the long-range  ambiguities in a definite way at an explicitly
or implicitly given scale while considering the Wilson coefficients
times (unknown) $B$-factors as phenomenological constants.

   As the above-mentioned perturbative QCD calculations in NLO are
believed to be reliable down to a scale of $O(1\,\mbox{GeV})$
(in \cite{buras3,buras4} the charm quark mass $m_c= 1.3$ GeV is used
as a matching scale), it is very desirable, however, that the
effective scale for the long-range calculation reaches this value
too.
   Therefore, our strategy includes a consistent calculation of the
matrix elements up to $O(p^6)$, and furthermore some consideration of
the effects from vector, axial-vector and scalar resonances which can
be included by the procedure of reduction \cite{bijnens1,reduction},
i.e. a certain recalculation of structure coefficients of the
effective chiral lagrangians involved.

   In this note we reconsider in the above manner especially the
$K\to 2\pi$ channels with matrix elements of four-quark operators
calculated at $O(p^6)$ in momentum expansions within the chiral
lagrangian approach.
   This approach is based on the bosonized version of the weak
lagrangian (\ref{weak-lagr}) and chiral effective meson lagrangians
of higher orders with the structure coefficients fixed by bosonization of
the four-quark NJL-type interaction.
   According to Weinberg's power-counting scheme \cite{weinberg}, the
calculation involves tree-level, one- and two-loop diagrams.
   The method of superpropagator regularization \cite{sp-volkov} was
used to fix UV divergences arising at the loop level.
      The aim of this note is to give the first numerical results of
this calculation concerning $K\to 2\pi$ amplitudes as they lead to
new estimates of $\varepsilon^{'}/\varepsilon$ (see section~3).
   In section~2 and appendices A and B we repeat all relevant
definitions taken from earlier work.

%============================================================================
%\begin{center}
%\large {\bf 2. Lagrangians and currents}
%\end{center}
\section{Lagrangians and currents}
%============================================================================

   In the present paper we use the operators ${\cal O}_i$
in the representation given in \cite{vzsh,bijnens-wise}:
\begin{eqnarray*}
{\cal O}_1 &=&   \bar{u}_L \gamma_\mu u_L \; \bar{d}_L \gamma^\mu s_L
       -   \bar{d}_L \gamma_\mu u_L \; \bar{u}_L \gamma^\mu s_L\,,
\nonumber\\
{\cal O}_2 &=&   \bar{u}_L \gamma_\mu u_L \; \bar{d}_L \gamma^\mu s_L
       +   \bar{d}_L \gamma_\mu u_L \; \bar{u}_L \gamma^\mu s_L
       + 2 \bar{d}_L \gamma_\mu d_L \; \bar{d}_L \gamma^\mu s_L
       + 2 \bar{s}_L \gamma_\mu s_L \; \bar{d}_L \gamma^\mu s_L\,,
\nonumber\\
{\cal O}_3 &=&   \bar{u}_L \gamma_\mu u_L \; \bar{d}_L \gamma^\mu s_L
       +   \bar{d}_L \gamma_\mu u_L \; \bar{u}_L \gamma^\mu s_L
       + 2 \bar{d}_L \gamma_\mu d_L \; \bar{d}_L \gamma^\mu s_L
       - 3 \bar{s}_L \gamma_\mu s_L \; \bar{d}_L \gamma^\mu s_L\,,
\nonumber\\
{\cal O}_4 &=&   \bar{u}_L \gamma_\mu u_L \; \bar{d}_L \gamma^\mu s_L
       +   \bar{d}_L \gamma_\mu u_L \; \bar{u}_L \gamma^\mu s_L
       -   \bar{d}_L \gamma_\mu d_L \; \bar{d}_L \gamma^\mu s_L\,,
\nonumber\\
{\cal O}_5 &=& \bar{d}_L \gamma_\mu \lambda^a_c s_L
  \left(\sum_{q=u,d,s}\bar{q}_R\,\gamma^\mu\,\lambda^a_c\,q_R\right)\,,
\quad
{\cal O}_6 = \bar{d}_L \gamma_\mu s_L
  \left( \sum_{q=u,d,s} \bar{q}_R \, \gamma^\mu \, q_R \right)\,,
\nonumber\\
{\cal O}_7 &=& 6\bar{d}_L \gamma_\mu s_L
  \left( \sum_{q=u,d,s} \bar{q}_R \, \gamma^\mu \, Q \, q_R \right)\,,
\quad
{\cal O}_8 = 6\bar{d}_L \gamma_\mu \lambda^a_c s_L
  \left(\sum_{q=u,d,s}\bar{q}_R\,\gamma^\mu\,\lambda^a_c\,Q\,q_R\right)\,,
\end{eqnarray*}
where $q_{L,R} = \frac12 \, (1\mp\gamma_5) q$; $\lambda^a_c$ are
the generators of the $SU(N_c)$ color group; $Q$ is the matrix of
electric quark charges.
   The operators ${\cal O}_{5,6}$ containing right-handed
currents are generated by gluonic penguin diagrams and the analogous
operators ${\cal O}_{7,8}$ arise from electromagnetic
penguin diagrams.
   The operators ${\cal O}_{1,2,3,5,6}$ and ${\cal O}_4$ describe the
transitions with $\Delta I = 1/2$ and $\Delta I = 3/2$, respectively,
while the operators ${\cal O}_{7,8}$ contribute to the transition with both
$\Delta I = 1/2$ and $\Delta I = 3/2$.

   The Wilson coefficients $C_i$ of the effective weak lagrangian
(\ref{weak-lagr}) with four-quark operators ${\cal O}_i$
are connected with the Wilson coefficients $c_i$ corresponding to the
basis of four-quark operators $Q_{i}$ given in Refs.
\cite{buras3,buras1}, by the following linear relations:
\begin{eqnarray}
&&C_1 = c_1-c_2+c_3-c_4+c_9-c_{10}\,,\quad
  C_2 = \frac{1}{5}\big(c_1+c_2-c_9-c_{10}\big)+c_3+c_4\,,
\nonumber \\&&
  C_3 = \frac{1}{5} C_4
      = \frac{1}{5}\bigg(
        \frac{2}{3}\big(c_1+c_2\big)+c_9+c_{10}\bigg)\,,
\nonumber \\&&
  C_5 = c_6\,,\quad C_6 = 2\bigg(c_5+\frac{1}{3}c_6\bigg)\,,\quad
  C_7 = \frac{1}{2}\bigg(c_7+2c_8\bigg)\,,\quad C_8 = \frac{1}{4}c_8\,.
\label{ci-param}
\end{eqnarray}

   The bosonized version of the effective Lagrangian (\ref{weak-lagr}) can be
expressed in the form \cite{bos-weak1,bos-weak2}:
 %%%%%%%%%%
 \def\jim{(J^1_{L\mu}\! - i J^2_{L\mu})}
 \def\rim{(J^1_R\! - i J^2_R)}
 \def\jiii{(J^3_{L\mu}\!+\! \frac1{\sqrt3} J^8_{L\mu})}
 \def\riii{(J^3_R\! -\!\frac1{\sqrt3} J^8_R\! - \sqrt\frac23 \, J^0_R)}
 \def\jivp{(J^4_{L\mu}\! + i J^5_{L\mu})}
 \def\livp{(J^4_L\! + i J^5_L)}
 \def\jvip{(J^6_{L\mu}\! + i J^7_{L\mu})}
 \def\lvip{(J^6_L\! + i J^7_L)}
 \def\rvip{(J^6_R\! + i J^7_R)}
 %%%%%%%%%%
\begin{eqnarray}
 {\cal L}^{mes}_{w} &=& \widetilde{G}_F\Bigg\{
  (-\xi_1\! + \xi_2\! + \xi_3) \bigg[ \jim \jivp - \jiii \jvip \bigg]
\nonumber\\
&&+ (\xi_1 + 5\,\xi_2) \sqrt\frac23 \, J^0_{L\mu} \jvip
+ {10 \over \sqrt3} \, \xi_3 \, J^8_{L\mu} \jvip
\nonumber\\
&&+ \xi_4 \bigg[ \jim \jivp + 2 \, J^3_{L\mu} \jvip \bigg]
\nonumber\\
&&- 4 \, \xi_5 \bigg[ \rim \livp - \riii \lvip
\nonumber\\
&&\qquad - \sqrt\frac23 \, \rvip (\sqrt2 J^8_L - J^0_L)
           \bigg]
\nonumber\\
&&+ \xi_6 \, \sqrt\frac32 \, \jivp J^0_R
+ 6 \, \xi_7 \, \jvip (J^3_{R\mu} + \frac1{\sqrt3} \, J^8_{R\mu})
\nonumber\\
&&- 16 \, \xi_8 \, \bigg[ \rim \livp + \frac12 \, \riii \lvip
\nonumber\\
&&\qquad + \frac1{\sqrt6} \, \rvip (\sqrt2 \, J^8_L - J^0_L)
            \bigg]
 \Bigg\} + \mbox{h.c.}
\label{weak-mes}
\end{eqnarray}
   Here $\widetilde{G}_F = \sqrt2\,G_F\,V_{ud}V^{*}_{us}$,
$J^a_{L/R \, \mu}$ and $J^a_{L/R}$ are bosonized $(V\mp A)$
and $(S\mp P)$ meson currents corresponding to the quark
currents
$\bar q \gamma_\mu \frac14 (1\mp \gamma^5) \lambda^a q$
and densities
$\bar q \frac14 (1\mp \gamma^5) \lambda^a q$,
respectively ($\lambda^a$ are the generators of the $U(3)_F$ flavor group);
\begin{eqnarray}
&&\xi_1 = C_1 \left( 1 - {1 \over N_c} \right)\,, \qquad
\xi_{2,3,4} = C_{2,3,4} \left( 1 + {1 \over N_c} \right)\,,
\nonumber\\
&&\xi_{5,8} = C_{5,8}\left( 1 - {1 \over N_c^2} \right) + {1 \over 2 N_c} C_{6,7}\,, \qquad
\xi_{6,7} = C_{6,7}\,,
\label{xi-param}
\end{eqnarray}
where the color factor ${1/N_c}$ originates from the Fierz-transformed
contribution to the nonleptonic weak effective chiral Lagrangian
\cite{bos-weak1,bos-weak2}.

   Only the even-intrinsic-parity sector of the chiral lagrangian is required
to describe nonleptonic kaon decays up to and including $O(p^6)$.
   In the bosonization approach this sector is obtained from the
modulus of the logarithm of the quark determinant of the NJL-type
models \cite{njl} using the path-integral technique (see
\cite{bos-strong} and references therein).
   The meson currents $J^a_{L/R\mu}$ and $J^a_{L/R}$ are obtained from this
quark determinant by variation over additional external sources
associated with corresponding quark currents and densities
\cite{bos-weak1,bos-weak2}.
   From the momentum expansion of the quark determinant to $O(p^{2n})$
one can derive the strong lagrangian for mesons ${\cal L}_{eff}$ of
the same order and the corresponding currents $J^a_{L/R\mu}$ and
$J^a_{L/R}$ to the order $O(p^{2n-1})$ and $O(p^{2n-2})$, respectively.

   For example, from the terms of quark determinant of $O(p^2)$ one gets
\begin{eqnarray}
&&{\cal L}^{(p^2)}_{eff}=-\frac{F^2_0}{4} \,\mbox{tr}\,
                      \big( L^2_{\mu}\big)
                     +\frac{F_0^2}{4} \,\mbox{tr}\,
                      \big( \chi U^\dagger + U\chi^\dagger \big)\,,
\nonumber \\&&
   J^{(p^1)a}_{L\mu} = \frac{iF^2_0}{4}\,\mbox{tr}\,(\lambda^a L_\mu)\,,
\quad
   J^{(p^0)a}_{L} =
   \frac{F^2_0}{4}\overline{m}R\,\,\mbox{tr}\,(\lambda^a U)\,,
\label{l2}
\end{eqnarray}
where $U = \exp \left(\frac{i\sqrt{2}}{F_0} \varphi \right)$,
with $\varphi$ being the pseudoscalar meson matrix,
and $L_{\mu}=D_{\mu}U\,U^\dagger$,
$D_{\mu}U = \partial_{\mu}U + (A^L_{\mu}U - U A^R_{\mu})$ and
$A^{R/L}_\mu = V_\mu \pm A_\mu$ are right/left-handed combinations of
vector and axial-vector fields.
   Furthermore, $F_0 \approx 90$ MeV is the bare coupling constant of pion
decay, $\chi =\mbox{diag}\big(\chi^2_u,\chi^2_d,\chi^2_s\big)
= -2m_0\!\!<\!\!\overline{q}q\!\!>\!\!F_0^{-2}$ is the meson mass
matrix, $\chi^2_u = 0.0114\,\mbox{GeV}^2$, $\chi^2_d = 0.025\,\mbox{GeV}^2$, 
$\chi^2_s = 0.47\,\mbox{GeV}^2$, $m_0$ is the current quark mass matrix,
$<\!\!\overline{q}q\!\!>$ is the quark condensate, $\overline{m} \approx 265$
MeV is an average constituent quark mass, and
$R=<\!\!\overline{q}q\!\!>\!\!/(\overline{m}F_0^2)$.

   At $O(p^4)$ one gets
\begin{eqnarray}
{\cal L}^{(p^4)}_{eff}&\Rightarrow&
    \bigg(L_1-\frac{1}{2}L_2 \bigg)\,\big( \,\mbox{tr}\, L^2_{\mu}\big)^2
  + L_2 \,\mbox{tr}\, \bigg(\frac{1}{2}[L_{\mu},L_{\nu}]^2
  + 3(L^2_{\mu})^2\bigg)
  + L_3 \,\mbox{tr}\, \big[ (L^2_{\mu})^2 \big]
\nonumber \\&&
  - L_4 \,\mbox{tr}\, \big( L^2_{\mu}\big)\,
        \,\mbox{tr}\, \big( \chi U^\dagger + U \chi^\dagger \big)
  - L_5 \,\mbox{tr}\,\Big( L^2_{\mu}
          \big( \chi U^\dagger + U \chi^\dagger \big)\Big)
\nonumber \\&&
  + L_8 \,\mbox{tr}\, \Big( (\chi ^\dagger U)^2 + (\chi U^\dagger)^2 \Big)
  + H_2 \,\mbox{tr}\, \chi \chi ^\dagger \,\,,
\nonumber \\
   J^{(p^3)a}_{L\mu} &\Rightarrow& i\,\mbox{tr}\,\bigg\{\lambda^a\bigg[
    L_4 L_\mu\,\mbox{tr}(\chi U^\dagger +U\chi^\dagger )
  + \frac{1}{2}L_5 \{ L_\mu ,(\chi U^\dagger +U\chi^\dagger ) \}\bigg]
    \bigg\}\,,
\nonumber \\
   J^{(p^2)a}_{L} &\Rightarrow& -\overline{m}R\,\mbox{tr}\,
                                 \Big\{\lambda^a \Big[
    L_4 U\,\mbox{tr}(L^2_\mu)
   +L_5 (L^2_\mu U) -2L_8 U\chi^\dagger U -H_2\chi \Big]\Big\}\,,
\label{l4}
\end{eqnarray}
where $L_i$ and $H_2$ are structure constants introduced by Gasser and
Leutwyler \cite{gasser}.
   For the sake of brevity, here and in following expressions for the
lagrangian and currents generated at $O(p^6)$, given in appendix A, we
restrict ourselves to the terms which are necessary to calculate the decay
$K\to 2\pi$ at $O(p^6)$.
   We do not show explicitly the terms of the effective action at $O(p^8)$
generating the scalar current $J^{(p^6)}_L$ which is necessary for the
full calculation of the tree-level matrix elements at $O(p^6)$ for the
penguin operators, since the corresponding contributions turn out to
be negligibly small.

    The structure constants $L_i$, $H_i$ and $Q_i$ should be obtained
from the modulus of the logarithm of
quark determinant of the NJL-type model which explicitly contains,
apart from the pseudoscalar Goldstone bosons, also scalar, vector and
axial-vector resonances as dynamic degrees of freedom.
    However, in order to avoid double counting in calculating
pseudoscalar meson amplitudes when taking into account resonance
degrees of freedom, one has to integrate out (reduce) these resonances
in the generating functional of the bosonization approach.
    As a consequence of this procedure, the structure coefficients of
pseudoscalar low-energy interactions will be quite strongly modified.
In this way one effectively takes into account resonance-exchange
contributions \cite{bijnens1,reduction}.
   The explicit expressions for the structure constants of
non-reduced and reduced effective meson lagrangians are given in
appendix B.

   In the context of the power counting rules we have to give some
comments concerning the additional symmetry breaking term
$\sim\mbox{tr}(m_0D^2U)$ which appears in
\cite{buras2,paschos,belkov-cp}.
   This term leads to nonzero meson matrix element of the gluonic
penguin operator due to the appearance of an additional contribution to
the scalar density $\sim\mbox{tr}(D^2U)$.
   In the bosonization approach it arises at $O(p^4)$ from that
term of the divergent part of the quark determinant which also generates
the kinetic term of $O(p^2)$:
\begin{eqnarray}
{\cal L}_{div} &\Rightarrow&
\frac{N_c}{16 \pi^2}\,y\,\mbox{tr} \Big[
          D^{\mu}(\overline{m} U+m_0)\,
\overline{D}_{\mu}(\overline{m} U+m_0)^\dagger\Big]
\nonumber \\&&
  = \frac{F^2_0}{4}\mbox{tr}
    \big( D^{\mu}U\,\overline{D}_{\mu}U^\dagger \big)
   -\frac{F^2_0}{4} \frac{1}{\Lambda^2_\chi}\mbox{tr}
    \big( \chi^\dagger D^2U + \chi \overline{D}^2U^\dagger \big)\,,
\label{ldiv}
\end{eqnarray}
 where
$\overline{D}_{\mu} U^\dagger = \partial_{\mu}U^\dagger
                              + (A^R_{\mu}U^\dagger - U^\dagger A^R_{\mu})$,
$y = 4\pi^2F_0^2/(N_c\overline{m}^2)$,
$\Lambda^2_\chi = \overline{m}^2/x$ with
$x = -\overline{m}F_0^2/(2\!\!<\!\!\overline{q}q\!\!>)$.
    For the phenomenological value of the quark condensate
$<\!\!\overline{q}q\!\!>^{1/3} \approx -220~\mbox{MeV}$ one obtains
$\Lambda_\chi \approx 839$ MeV (to be compared with
$\Lambda_\chi \approx 1020$ MeV in \cite{buras3,buras1}).
    The term $\sim\mbox{tr}(m_0D^2U)$ does not appear explicitly in the
effective lagrangian (\ref{l4}) because after transformations of
double derivatives according to the equations of motion
\begin{eqnarray*}
D^2U &=&
  -D_\mu U\,\overline{D}^{\mu}U^\dagger\cdot U
  -\frac{1}{2}\big(U\chi^\dagger U -\chi\big)
  -\frac{1}{6}U\,\mbox{tr}\,
   \big(\chi U^\dagger -U\chi^\dagger\big)\,,
\\
\overline{D}^2U^\dagger &=&
  -U^\dagger D_\mu U\,\overline{D}^{\mu}U^\dagger
    - \frac{1}{2}\big(U^\dagger \chi U^\dagger -\chi^\dagger\big)
    + \frac{1}{6}U^\dagger\,\mbox{tr}\,
      \big(\chi U^\dagger -U\chi^\dagger\big)\,,
\end{eqnarray*}
it is transformed into a combination of terms contributing to the structure
coefficients $L_5$, $L_8$ and $H_2$ (compare Eq.~(\ref{l4})):
\begin{eqnarray}
-\frac{F^2_0}{4} \frac{1}{\Lambda^2_\chi}\mbox{tr}
\big(\chi^\dagger D^2U+\chi \overline{D}^2U^\dagger\big) &=&
\frac{N_c}{16\pi^2}\bigg[
  - xy\,\mbox{tr}\,\Big( L^2_{\mu}
         \big( \chi U^\dagger + U \chi^\dagger \big)\Big)
\nonumber \\&&
  +\frac{1}{2}xy\,\mbox{tr}\,\Big(( \chi^\dagger U)^2
                                   +\chi U^\dagger)^2 \Big)
  - xy\,\mbox{tr}\, \chi \chi ^\dagger
\nonumber \\&&
  - \frac{1}{6}xy\,\Big(\mbox{tr}\,\big( \chi U^\dagger
                                        -U\chi^\dagger\big)
                   \Big)^2\bigg]\,.
\label{D2-transformed}
\end{eqnarray}
   The last term in Eq.~(\ref{D2-transformed}) does not contribute in
$K\to 2\pi$ decays, therefore the corresponding term ($\sim L_7$) in
(\ref{l4}) is dropped.
    The corresponding numerical contributions to the structure
coefficients are $L_5 = 2.9\cdot 10^{-3}$, $L_8 = 1.4\cdot 10^{-3}$
and $H_2 =-2.9\cdot 10^{-3}$.
   The complete expressions (\ref{lhcoeff}) in appendix B contain
further contributions to $L_5$, $L_8$ and $H_2$, which originate from
the divergent as well as the finite parts of the quark determinant
at $O(p^4)$, and result in the numerical values
$L_5 = 0.98\cdot 10^{-3}$, $L_8 = 0.36\cdot 10^{-3}$ and
$H_2 = 1.01\cdot 10^{-3}$ (see table 6 in appendix B).

%============================================================================
%\begin{center}
%\large {\bf 3. Amplitudes and phenomenological results}
%\end{center}
\section{Amplitudes and phenomenological results}
%============================================================================

   Using isospin relations, the $K \to 2\pi$ decay amplitudes can be
parameterized  as
\begin{eqnarray*}
T_{K^+\to\pi^+\pi^0} &=& {\sqrt3\over 2}\,A_2\,,
\nonumber \\
T_{K^0_S\to\pi^+\pi^-} &=& \sqrt{2\over 3}\,A_0+{1\over\sqrt3}\,A_2\,,
\qquad
T_{K^0_S\to\pi^0\pi^0} = \sqrt{2\over 3}\,A_0-{2\over\sqrt3}\,A_2\,.
\end{eqnarray*}
   The isotopic amplitudes $A_{2,0}$ determine the $K \to 2\pi$
transitions into states with isospin $I=2,0$, respectively:
$$
A_2 = a_2 \, e^{i\delta_2}\,, \qquad A_0 = a_0 \, e^{i\delta_0}\,,
$$
where $\delta_{2,0}$ are the phases of $\pi\pi$-scattering.
   It is well known that direct $CP$ violation results in an additional
(small) relative phase between $a_2$ and $a_0$.
   Let us next introduce the contributions of the four-quark
operators ${\cal O}_i$ to the isotopic amplitudes ${\cal A}_I^{(i)}$
by the relations
\begin{eqnarray}
A_I&=&{\cal F}_I {\cal A}_I\,,\quad
{\cal A}_I = -i\,\sum_{i=1}^8\xi_i {\cal A}_I^{(i)}\,,
\label{defa}
\end{eqnarray}
where
${\cal F}_2 =\sqrt2{\cal F}_0
            = {\sqrt3\over2}\widetilde{G}_F F_0(m_K^2-m_\pi^2)$.

   At $O(p^2)$ we obtain for the nonzero tree-level amplitudes
${\cal A}_I^{(i)}$ the following expressions:
\begin{eqnarray}
&&\!\!\!\!\!\!\!\!\!\!
{\cal A}_0^{(1)}=-{\cal A}_0^{(2,3)}=-1=-{\cal A}_0^{(4)}\,,\quad
{\cal A}_0^{(7)} = -{\cal A}_2^{(7)} = 2\,,\quad
{\cal A}_0^{(5)} = -32\bigg(\frac{R\overline{m}}{F_0}\bigg)^2 L_5 \,,
\nonumber \\&&\!\!\!\!\!\!\!\!\!\!
{\cal A}_0^{(8)} = \frac{16(R\overline{m})^2}{m^2_K-m^2_\pi}
\bigg\{1-\frac{2}{F^2_0}\bigg[ 6L_4(\chi^2_s+\chi^2_d+\chi^2_u)
\nonumber \\&&\quad\quad\quad\quad\quad\quad\quad\quad\quad\quad
        + (L_5-4L_8)(\chi^2_s+3\chi^2_d+2\chi^2_u)
        + 2L_5 m_\pi^2 \bigg]\bigg\}\,,
\nonumber \\&&\!\!\!\!\!\!\!\!\!\!
{\cal A}_2^{(8)} = \frac{8(R\overline{m})^2}{m^2_K-m^2_\pi}
\bigg\{1-\frac{2}{F^2_0}\bigg[ 6 L_4(\chi^2_s+\chi^2_d+\chi^2_u)
\nonumber \\&&\quad\quad\quad\quad\quad\quad\quad\quad\quad\quad
        + (L_5-4L_8)(\chi^2_s+3\chi^2_d+2\chi^2_u)
        + 2L_5 m_K^2 \bigg]\bigg\}\,.
\label{A02-p2}
\end{eqnarray}
   The $L_8$ and $H_2$ contributions in the penguin operators
${\cal O}_{5,8}$ also have a tadpole contribution from
$K \to (\mbox{vacuum})$ included through strong rescattering,
$K\to \pi\pi K$ with $K \to (\mbox{vacuum})$.
   At $O(p^2)$, in case of the penguin operator ${\cal O}_5$, the $L_8$
and $H_2$ contributions in the direct matrix element from $K\to 2\pi$
vertices, are fully cancelled by the tadpole diagrams
\footnote{We thank W.A.~Bardeen and A.J.~Buras for drawing our attention
          to this point.}.
   This is due to the possibility to absorb the tadpole contribution
into a redefinition of the $K\to 2\pi$ vertex if all particles are on
mass shell.
  In higher orders, however, such a cancellation is generally not
expected.
  Therefore, one of the main aims of the present paper is to perform all
calculations up to $O(p^6)$ to clarify this point quantitatively.

   Some interesting observations on the difference of the momentum behavior
of penguin and nonpenguin operators can be drawn from power-counting
arguments.
   According to Eq.\ (\ref{l2}) the leading contributions to the
vector and scalar currents are of $O(p^1)$ and $O(p^0)$, respectively.
   Since in our approach the nonpenguin operators are constructed out of
the products of $(V-A)$-currents $J^a_{L\mu}$, while the penguin operators
are products of $(S-P)$-currents $J^a_L$, the lowest-order contributions
of nonpenguin and penguin operators are of $O(p^2)$ and $O(p^0)$,
respectively.
   However, due to the well-known cancellation of the contribution of
gluonic penguin operator ${\cal O}_5$ at the lowest order
\cite{chivukula}, the leading gluonic penguin as well as nonpenguin
contributions start from $O(p^2)$
\footnote{There is no cancellation of the contribution of the
          electromagnetic penguin operator ${\cal O}_8$ at the lowest
          order and the first terms in the expressions (\ref{A02-p2})
          for ${\cal A}_{0,2}^{(8)}$ correspond to the contributions
          at $O(p^0)$.}.
   Consequently, in order to derive the $(V-A)$-currents which
   contribute to the
nonpenguin transition operators at leading order, it is sufficient to
use the terms of the quark determinant to $O(p^2)$ only.
   At the same time the terms of the quark determinant to $O(p^4)$
have to be
kept for calculating the penguin contribution at $O(p^2)$ since it
arises from the combination of $(S-P)$-currents from Eqs.\ (\ref{l2}) and
(\ref{l4}), which are of $O(p^0)$ and $O(p^2)$, respectively.
   In this subtle way a difference in momentum behaviour is revealed
between matrix elements for these two types of weak transition operators;
it manifests itself more drastically in higher-order lagrangians and
currents.
    This fact makes penguins especially sensitive to higher order effects.

    Using the truncated lagrangian (\ref{ldiv}) only, the gluonic penguin
operator matrix element at $O(p^2)$ is
\begin{equation}
{\cal A}_0^{(5)} = 4R \approx -20.0\,.
\label{peng1}
\end{equation}
    However,  taking into account the additional contribution from other
parts of the quark determinant at $O(p^4)$, we get a suppression of
this matrix element which after substitution of the full expression
for the non-reduced structure constant $L_5$ from Eq.~(\ref{lhcoeff}) becomes
$$
{\cal A}_0^{(5)} = 4R \bigg(1-\frac{1}{y}\bigg)
                 \approx -6.8\,.
$$
   The modification of the structure constant after reduction of resonances
($L_5^{red} = 1.64\cdot 10^{-3}$, see appendix B) leads to an increase in
absolute value of the gluonic penguin matrix element to
${\cal A}_0^{(5)} \approx -11.2$, which, however, is still about factor 2
smaller than (\ref{peng1}).

   Table 1 presents the modification of the amplitudes
${\cal A}_I^{(i)}$ when including successively the higher order
corrections $O(p^4)$ and $O(p^6)$ and the reduction of meson resonances.
   Our calculations involve Born and one- and two-loop meson diagrams
and take into account isotopic symmetry breaking
($\pi^0 - \eta -\eta^{'}$ mixing).
   In our approach, the UV divergences resulting from meson loops at
$O(p^4)$ and $O(p^6)$ were separated using the superpropagator
regularization method \cite{sp-volkov} which is particularly
well suited to the treatment of loops in nonlinear chiral theories.
   The result is equivalent to the dimensional regularization
technique, the difference being that the scale parameter $\mu$ is no
longer arbitrary but fixed by the inherent scale of the chiral theory
$\tilde{\mu}=4\pi F_0\approx 1$ GeV, and the UV divergences have to be
replaced by a finite term using the substitution
$$
\big({\cal C}-1/\varepsilon\big) \to C_{SP} = -1+4{\cal C} +\beta \pi\,,
$$
where ${\cal C} = 0.577$ is Euler's constant,
$\varepsilon = (4-D)/2$ and $\beta$ is an arbitrary constant
introduced by the Sommerfeld-Watson integral representation of the
superpropagator.
   The splitting of the decay constants $F_\pi$ and $F_K$ is used at
$O(p^4)$ to fix $C_{SP}\approx 3.0$.

   The strong interaction phases $\delta_{2,0}$ arise first at
$O(p^4)$, but for the quantitative description of the
phases it is necessary to go beyond $O(p^4)$.
   At $O(p^4)$, for the $\pi\pi$-scattering phase shifts and their
difference $\Delta=\delta_0-\delta_2$, we have obtained the values
of $\delta_0 \approx 22^\circ$, $\delta_2 \approx -13^\circ$,
$\Delta \approx 35^\circ$ which are in agreement with \cite{kambor}.
   At $O(p^6)$, we have obtained $\delta_0 \approx 35^\circ$,
$\delta_2 \approx -9^\circ$, $\Delta \approx 44^\circ$, in better
agreement with the experimental value $\Delta^{exp} = (48 \pm 4)^\circ$
\cite{belkov-kostyukhin}.

   In our approach the parameters $\xi_i$ in Eq.~(\ref{defa}) are
treated as phenomenological ($\mu$-independent) parameters to be fixed
from the experimental data.
   They can be related to the $\mu$-dependent QCD predicted
$\xi_i(\mu)$ with the help of some $\mu$-dependent $B_i$-factor defined
as
$$
   \xi_i^{ph}=\xi_i(\mu)B_i(\mu).
$$
   Table 2 shows the QCD predictions for the coefficients
$\xi_i(\mu) = \xi_i^{(z)}(\mu)+\tau \xi_i^{(y)}(\mu)$ which correspond
to the Wilson coefficients
$$
  c_i(\mu) = z_i(\mu)+\tau y_i(\mu),\quad
  \tau = -\frac{V_{td}V_{ts}^{*}}{V_{ud}V_{us}^{*}}\,,
$$
from the table XVIII of Ref.~\cite{buras3} calculated numerically from
perturbative QCD at $\mu = 1$ GeV for $m_t = 170$ GeV in leading (LO)
and next-to-leading order in different renormalization schemes (NDR
and HV).
    $\xi_i^{(z)}$ and $\xi_i^{(y)}$ were obtained from $z_i$
and $y_i$, respectively, using the Eqs.~(\ref{ci-param}) and
(\ref{xi-param}).

   As we cannot calculate the factors $B_i(\mu)$ theoretically, they
can only be fixed from data in the spirit of the semi-phenomenological
approach \cite{buras3,buras1,buras4}.
   Table 1 shows that the amplitudes of $K\to 2\pi$ decays
are dominated by the contribution of the operators ${\cal O}_i$ with
$i =1,2,3,4,5,8$.
   Moreover, in case of the operators ${\cal O}_{1,2,3}$, the first
term in the combination $(-\xi_1\! + \xi_2\! + \xi_3)$ dominates in
the effective weak meson lagrangian (\ref{weak-mes}).
   Thus, the isotopic amplitudes can be given in the approximation
of the dominating contributions of four-quark operators as
\begin{eqnarray}
{\cal A}_I &=& {\cal A}_I^{(z)}+\tau{\cal A}_I^{(y)}\,,
\nonumber \\
{\cal A}_I^{(z,y)}&=&
    \Big[-\xi_1^{(z,y)}(\mu)+\xi_2^{(z,y)}(\mu)+\xi_3^{(z,y)}(\mu)\Big]
     B_1(\mu)\,{\cal A}_I^{(1)}
   +\xi_4^{(z,y)}(\mu)B_4(\mu)\,{\cal A}_I^{(4)}
\nonumber \\ &&
   +\xi_5^{(z,y)}(\mu)B_5(\mu)\,{\cal A}_I^{(5)}
   +\xi_8^{(z,y)}(\mu)B_8(\mu)\,{\cal A}_I^{(8)}
    \Big]\,,
\label{AI-approximation}
\end{eqnarray}
and
$$
A_I = \big( a_I^{(z)} +\tau a_I^{(y)}\big)\, e^{i\delta_I}\,.
$$
   At least two  factors $B_1$ and $B_4$ can be estimated from the
experimental values ${\cal A}_0^{exp} \approx 10.9$ and
${\cal A}_2^{exp} \approx 0.347$ while the other two (penguin) factors
$B_5$ and $B_8$ should be fixed from other data or restricted by
theoretical arguments.

   The parameter $\varepsilon^{'}$ of direct $CP$-violation in
$K\to2\pi$ decays can be expressed by the formulae
$$
\varepsilon^{'} =
-{\omega\over\sqrt2}\,{\mbox{Im}\,a_0\over\mbox{Re}\,a_0}
  \big(1-\Omega\big)\,\mbox{e}^{i(\pi/2+\delta_2-\delta_0)}\,,\quad
\omega = \frac{\mbox{Re}\,a_2}{\mbox{Re}\,a_0}\,,\quad
\Omega = \frac{1}{\omega}\frac{\mbox{Im}\,a_2}{\mbox{Im}\,a_0}\,,
$$
and the ratio $\varepsilon^{'}/\varepsilon$ can be estimated as
\begin{equation}
\frac{\varepsilon^{'}}{\varepsilon} =
 \mbox{Im}\,\lambda_t\,\big(P_0-P_2),\quad
  P_I= \frac{\omega}{\sqrt{2}\varepsilon |V_{ud}||V_{us}|}\,
       \frac{a_I^{(y)}}{a_I^{(z)}}\,,
\label{P0P2}
\end{equation}
with
$\mbox{Im}\,\lambda_t = \mbox{Im}\,V^{*}_{ts}V_{td}
                    = |V_{ts}||V_{td}| \mbox{sin} \delta$
in the standard parameterization of the CKM matrix.
    In our estimates for $\varepsilon^{'}/\varepsilon$ we will use the
restriction \cite{buras3,buras4}
\begin{equation}
   0.86\cdot10^{-4} \le \mbox{Im}\,\lambda_t \le 1.71\cdot 10^{-4}
\label{lambda}
\end{equation}
obtained from the phenomenological analysis of indirect $CP$ violation
in $K\to 2\pi$ decay and $B^0-\overline B^0$ mixing.

   Table 3 gives the estimates of $\varepsilon^{'}/\varepsilon$ from a
semi-phenomenological approach obtained after fixing the correction factors
$B_1$ and $B_4$ for isotopic amplitudes in the representation
(\ref{AI-approximation}) by experimental ($CP$-conserving) data on
$\mbox{Re}\,{\cal A}_{0,2}$, and setting $B_5 = B_8 =1$.
   We have used the matrix elements of the operators ${\cal O}_i$ displayed
in table 1, and the theoretical values $\xi_i(\mu)$ from table 2.
   Taking into account the dependence of the results on the
renormalization scheme, we have obtained without reducing resonances
(see table 1a) the following upper and lower bounds for
$\varepsilon^{'}/\varepsilon$, corresponding to the interval
(\ref{lambda}) for $\mbox{Im}\,\lambda_t$:
$$
-4.7\cdot10^{-4}\le\varepsilon^{'}/\varepsilon\le -0.4\cdot 10^{-4}\,.
$$
   The peculiarity of these results lies in the observation that all
estimates of $\varepsilon^{'}/\varepsilon$ lead to negative values.
   This is related to the fact that in the case corresponding to table 1a
the contribution of gluonic penguins to $\Delta I = 1/2$ transitions appears
to be suppressed, leading, after the interplay between gluonic and
electromagnetic penguins, to the relation $P_0 < P_2$ for the two competing
terms in (\ref{P0P2}).
   Generally speaking, $\Delta I = 1/2$ transitions loose importance
compared to $\Delta I = 3/2$ when estimating $\varepsilon^{'}/\varepsilon$.
    The situation changes after the reduction of resonances, due to
a relative enhancement of the matrix element for the operator ${\cal O}_5$
(see table 1b).
    The bounds obtained in this case are
$$
-3.0\cdot10^{-4}\le\varepsilon^{'}/\varepsilon\le 3.6\cdot 10^{-4}\,,
$$
that substantially agrees with the bounds given by
\cite{buras3,buras4}.

    So, our calculations have shown that especially the penguin matrix
elements are most sensitive to various refinements: higher-order derivative
terms in chiral lagrangians, the reduction of meson resonances,
$\pi^0 - \eta - \eta^{'}$ mixing, meson loop corrections.
   It should be added that the modification of penguin matrix
elements, discussed in this note, is much more important for gluonic
than for electromagnetic penguin transitions.
   This is obvious from the observation that the latter at lowest
order contain terms of $O(p^0)$ which are left unchanged when taking
into account the additional terms derived from the quark determinant
at $O(p^4)$.

    Finally, we give some results concerning the dependence of the
above semi-phenomenological estimates for $\varepsilon^{'}/\varepsilon$ on
the choice of the penguin correction factors $B_5$ and $B_8$.
    In table 4 we present the $B_5$-dependence of
$\varepsilon^{'}/\varepsilon$ estimated with reducing meson
resonances for the central value
$\mbox{Im}\,\lambda_t = 1.29\cdot 10^{-4}$ and $B_8 =1$.
    The corresponding $B_8$-dependence of $\varepsilon^{'}/\varepsilon$
($B_5 =1$) is given in table 5.
    The sensitivity to a small ($\sim 5\%$) uncertainty from the 
phenomenological fixation of $<\!\!\overline{q}q\!\!>^{1/3}$ is also
shown in figures 1 and 2.
    It turns out to be problematic to obtain the value of 
$\varepsilon^{'}/\varepsilon$ as large as $(20-40)\cdot 10^{-4}$ and
consistent with the NA31 result \cite{NA31-CP} even varying the input 
parameters $<\!\!\overline{q}q\!\!>^{1/3}$, $B_5$ and $B_8$ in a wide 
region.
    Taking $B_5 = B_8 =1$, a value of 
$|<\!\!\overline{q}q\!\!>^{1/3}| > 300\,\mbox{MeV}$ would be required in
order to reach $\varepsilon^{'}/\varepsilon > 20\cdot 10^{-4}$, which
is far above its phenomenological limits derived within the usual
chiral framework.

    Since our results are very sensitive to the relative contribution
of the gluonic penguin operator, the question of its phenomenological
separation in $K\to 2\pi$ decays becomes critical, in the context
of the $\Delta I = 1/2$ rule as well as for the very important problem of
direct $CP$-violation, where at least one experimental result
\cite{NA31-CP} would lead to some revision of the present picture.
    $CP$-conserving $K\to 2\pi$ data alone are clearly not sufficient
for such a separation.
    It could be accomplished, on the other hand, when taking into account
Dalitz-plot data for $K\to 3\pi$ as well as differential distributions
for radiative decays $K\to 2\pi\gamma$, $K\to \pi 2\gamma$, which are
described by the same lagrangian (\ref{weak-lagr}).
     As emphasized above, the reason for this possibility is found in
the difference in momentum power counting behaviour between penguin
and non-penguin matrix elements, which appears in higher orders
of chiral theory, when calculating various parameters of differential
distributions, for instance, slope parameters of the Dalitz-plot for
$K\to 3\pi$.
    A substantial improvement in the accuracy of such experimental
data (mostly being of older dates) would be very helpful for such a
phenomenological improvement of the theoretical situation for
$\varepsilon^{'}/\varepsilon$ (see \cite{bos-weak2} for a discussion
of this point and \cite{hyperon,tnf} for some recent measurements).
    Of course, for all these model developments the new experiments at CERN,
FNAL and planned at Frascati have to be considered as crucial if
the accuracy level of $10^{-4}$ for $\varepsilon^{'}/\varepsilon$ is
obtained.

    The authors gratefully acknowledge fruitful and helpful discussions with
W.A. Bardeen and A.J.~Buras.

\vspace{5mm}
%============================================================================
\appendix
%\begin{center}
%\large {\bf Appendix A}
%\end{center}
\section*{Appendix A}
%============================================================================

   At $O(p^6)$ one gets
\footnote{The rather lengthy full expression for bosonized effective
lagrangian at $O(p^6)$ was presented in refs.\cite{p6-our,warsaw1}.}
\begin{eqnarray*}
{\cal L}_{eff}^{(p^6)} &\Rightarrow& \mbox{tr} \bigg\{
          Q_{12} \Big(\,\chi R^\mu U^\dagger \big(
            D_\mu D_\nu U + D_\nu D_\mu U \big) U^\dagger L^\nu
\nonumber \\ && \quad \quad
           +\chi^\dagger L^\mu U \big(
            \overline{D}_\mu \overline{D}_\nu U^\dagger
           +\overline{D}_\nu \overline{D}_\mu U^\dagger \big) UR^\nu
          \Big)
\nonumber \\ &&
         +Q_{13}\Big[
            \chi (  \overline{D}_\mu\overline{D}_\nu U^\dagger
                  L^\mu L^\nu
              + R^\nu R^\mu U
                \overline{D}_\mu \overline{D}_\nu U^\dagger )
\nonumber \\ && \quad \quad
           +\chi^\dagger (D_\mu D_\nu U R^\mu R^\nu
                         +L^\nu L^\mu D_\mu D_\nu U) \Big]
\nonumber \\ &&
         +Q_{14}\Big[{\chi}\Big(U^{\dagger}D_{\mu}D_{\nu}U\overline{D}^{\mu}
          \overline{D}^{\nu}U^\dagger+\overline{D}_{\mu}\overline{D}_{\nu}
          U^{\dagger}D^{\mu}D^{\nu}UU^{\dagger}\Big)
\nonumber \\ && \quad \quad
                 +{\chi}^{\dagger}\Big(U\overline{D}_{\mu}
          \overline{D}_{\nu}U^{\dagger}D^{\mu}D^{\nu}U
          +D_{\mu}D_{\nu}U\overline{D}^{\mu}
          \overline{D}^{\nu}U^{\dagger}U\Big)\Big]
\nonumber \\ &&
         +Q_{15} \chi^\dagger L_{\mu} \chi R^{\mu}
         +Q_{16}\Big( \chi^\dagger \chi R_{\mu} R^{\mu}
              +\chi \chi^\dagger L_{\mu} L^{\mu} \Big)
\nonumber \\ &&
         +Q_{17}\Big( U\chi^\dagger U \chi^\dagger L_{\mu}
              L^{\mu}+U^\dagger \chi  U^\dagger \chi R_{\mu} R^{\mu} \Big)
         +Q_{18}\Big[(\chi U^\dagger L_{\mu})^2
         + (\chi^\dagger U R_{\mu} )^2 \Big]
\nonumber \\ &&
         +Q_{19}\Big[(\chi  U^\dagger)^3
             + (\chi^\dagger U)^3 \Big]
         +Q_{20}\Big( U^\dagger \chi  \chi^\dagger \chi
                  +U \chi^\dagger \chi  \chi^\dagger \Big)
                         \bigg\}\,,
\label{l6}
\end{eqnarray*}
where $Q_i$ are structure constants introduced in \cite{warsaw1},
whereas $R_\mu =U^\dagger D_\mu U$.
   The corresponding terms of $(V\mp A)$ and $(S\mp P)$ bosonized
meson currents are given by
\begin{eqnarray*}
J^{(p^5)a}_{L \mu} &\Rightarrow&i\frac{1}{4} \mbox{tr} \bigg\{ \lambda^a \bigg[
 -2 Q_{14} \Big[
                (U\chi^{\dagger}+\chi U^{\dagger})D_{\mu}D_{\nu}U\,
                 U^{\dagger}L^{\nu}
               +D_{\mu}D_{\nu}U(U^{\dagger}\chi U^{\dagger}+\chi ^{\dagger})
                                   L^{\nu}
\nonumber \\&&~~~~~~~~~
               -U\overline{D}^{\nu}\Big(
                       (U^{\dagger}\chi +\chi ^{\dagger}U)\overline{D}_{\nu}
                                 \overline{D}_{\mu}U^{\dagger}
                        +\overline{D}_{\nu}\overline{D}_{\mu}U^{\dagger}
                          (U\chi ^{\dagger}+\chi U^{\dagger})
                                    \Big)
\nonumber \\&&~~~~~~~~~
               +L^{\nu}U\Big(
                       (U^{\dagger}\chi +\chi ^{\dagger}U)\overline{D}_{\mu}
                                 \overline{D}_{\nu}U^{\dagger}
                        +\overline{D}_{\mu}\overline{D}_{\nu}U^{\dagger}
                           (U\chi ^{\dagger}+\chi U^{\dagger})
                        \Big)
\nonumber \\&&~~~~~~~~~
               +D^{\nu}\Big((U\chi ^{\dagger}+\chi U^{\dagger})D_{\nu}D_{\mu}U
                        +D_{\nu}D_{\mu}U(U^{\dagger}\chi +\chi ^{\dagger}U)
                       \Big)U^{\dagger}
           \Big]
\nonumber \\&&
 +2 Q_{15} \big(U{\chi}^{\dagger}L_{\mu}{\chi}U^{\dagger}
                 +{\chi}U^{\dagger}L_{\mu}U{\chi}^{\dagger}
            \big)
 +2 Q_{16} \big(\big\{U{\chi}^{\dagger}{\chi}U^{\dagger},
                               L_{\mu}\big\}
                 +\big\{{\chi}{\chi}^{\dagger},L_{\mu}\big\}
           \big)
\nonumber \\&&
 +2 Q_{17} \big(\big\{\big(U{\chi}^{\dagger}\big)^2,L_{\mu}\big\}
                +\big\{\big({\chi}U^{\dagger}\big)^2,L_{\mu}\big\}
           \big)
\nonumber \\&&
 -4 Q_{18} \big( U{\chi}^{\dagger}L_{\mu}U{\chi}^{\dagger}
                +{\chi}U^{\dagger}L_{\mu}{\chi}U^{\dagger}
           \big)
           \bigg]\bigg\}\,,
\label{curv5}
\end{eqnarray*}
and
\begin{eqnarray*}
J^{(p^4)a}_L &\Rightarrow& \overline{m}R\mbox{tr}\Big\{ \lambda^a \Big[
  Q_{12}\, L^{\mu} U\{\overline{D}_{\mu},\overline{D}_{\nu}\}U^\dagger
           \, UR^{\nu}
 +Q_{13} \big( L^{\nu}L^{\mu}D_{\mu}D_{\nu}U
              +D_{\mu}D_{\nu}U\cdot R^{\mu}R^{\nu}\big)
\nonumber \\&&
 +Q_{14} \big( U\overline{D}^{\nu}\overline{D}^{\mu}U^{\dagger}
                             D_{\nu}D_{\mu}U
                            +D_{\nu}D_{\mu}U
                \overline{D}^{\nu}\overline{D}^{\mu}U^{\dagger}\cdot U\big)
\nonumber \\&&
 +Q_{15}\, L^{\mu} \chi R_{\mu}
 +Q_{16}\,\big( \chi R_{\mu}^2 +L_{\mu}^2 \chi \big)
 +Q_{17}  \big( U \chi^{\dagger} UR_{\mu}^2
          +L_{\mu}^2 U \chi^{\dagger} U \big)
\nonumber \\&&
 +2Q_{18}\, L^{\mu}U \chi^{\dagger} L_{\mu}U
 + Q_{19} \big(U\chi ^{\dagger}\big)^2U
 +Q_{20} (\chi U^{\dagger}\chi +\chi \chi ^{\dagger}U +U\chi ^{\dagger}\chi )
           \Big] \Big\}\,.
\label{curs4}
\end{eqnarray*}

\vspace{5mm}
%============================================================================
%\begin{center}
%\large {\bf Appendix B}
%\end{center}
\section*{Appendix B}
%============================================================================

    Without reduction of resonance degrees of freedom the structure
constants $L_i = N_c/(16 \pi^2)\cdot l_i$,
$H_2 = N_c/(16 \pi^2)\cdot h_i$ and
$Q_i = N_c/(32 \pi^2 \overline{m}^2)\cdot q_i$ are fixed from the bosonization
of an NJL-type model as
\begin{eqnarray}
&& l_1 = \frac{1}{2}l_2 = \frac{1}{24}\,,\quad
   l_3=-\frac{1}{6}\,,\quad
   l_4= 0 \,,\quad
   l_5= xy-x\,,
\nonumber \\
&& l_8= \frac{1}{2}xy-x^2y-\frac{1}{24}\,,\quad
   h_2=-xy-2x^2y+\frac{1}{12}
       +\frac{16\pi^2}{N_c}\frac{xF^2_0}{4\overline{m}^2}\,,
\label{lhcoeff}
\end{eqnarray}
and
\begin{eqnarray*}
&&
q_{12}= \frac{1}{60}\,,\quad
q_{13}= -\frac{1}{3}\bigg(\frac{1}{20}-x+c\bigg)\,,\quad
q_{14}= \frac{x}{6}\,,
\nonumber \\&&
q_{15}= \frac{2}{3}x\big(1-x\big)-
        \bigg(\frac{1}{3}-2x\bigg)c\,,\quad
q_{16}= -\frac{1}{120}+\frac{4}{3}x^2+\frac{x}{6}\big(1-4x\big)
        -2\bigg( x-\frac{1}{6} \bigg)c\,,
\nonumber \\&&
q_{17}=  \frac{1}{120}+\frac{x}{6}\big(1-4x\big)
        -\bigg(x+\frac{1}{6}\bigg)c\,,\quad
q_{18}= \frac{4}{3}x^2+\bigg(\frac{1}{6}-x\bigg)c\,,
\nonumber \\ &&
q_{19}=-\frac{1}{240}-x^2+\frac{2}{3}x^3
       +x\big(1+2xy\big)c\,,
\nonumber \\ &&
q_{20}= \frac{1}{240}+x^2+2\big(1-2y\big)x^3-x\big(1+2xy\big)c\,,
\label{qcoeff}
\end{eqnarray*}
where $x = -\overline{m} F_0^2/(2\!\!<\!\!\overline{q} q \! \! >)$,
$y = 4\pi^2F_0^2/(N_c\overline{m}^2)=1.5$ and $c=1-1/(6y)$.
    After reduction of the resonances, the structure coefficients
get the form
\begin{eqnarray*}
&&\!\!\!\!\!\!\!\!\!\!
l^{red}_1 =\frac{1}{2}l^{red}_2
          = \frac{1}{12}\bigg[ Z^8_A
          +2(Z^4_A-1)\bigg(\frac{1}{4}\tilde{y}(Z^4_A-1)-Z^4_A\bigg)\bigg]\,,
\nonumber \\ &&\!\!\!\!\!\!\!\!\!\!
l^{red}_3 = -\frac{1}{6}\bigg[ Z^8_A
          +3(Z^4_A-1)\bigg(\frac{1}{4}\tilde{y}(Z^4_A-1)-Z^4_A\bigg)\bigg]\,,
\nonumber \\ &&\!\!\!\!\!\!\!\!\!\!
l^{red}_4= 0 \,,\quad
l^{red}_5 = (\tilde{y}-1) \frac14 Z^6_A \,,\quad
l^{red}_8 = \frac{\tilde{y}}{16} Z^4_A \,,\quad
h^{red}_2 = \tilde{y} Z^2_A \bigg(\frac{Z^2_A}{8}-x\bigg)\,.
\label{lhred}
\end{eqnarray*}
and
\begin{eqnarray*}
&& \!\!\!\!\!\!\!\!\!\!
q^{red}_{12} = q^{red}_{13}= 0 \,,\quad
q^{red}_{14} = \frac{1}{24}Z^6_A\,,
\nonumber \\ &&\!\!\!\!\!\!\!\!\!\!
q^{red}_{16} = q^{red}_{17} =
-\frac{Z^6_A}{64}\bigg\{\tilde{y}\!
       -Z^2_A\bigg[4\!-6\Big(1+4(1\!-Z^2_A)\Big)(1\!-\tilde{y})
       +4\Big(1+16(1\!-Z^2_A)\Big)
         \frac{1\!-\tilde{y}}{\tilde{y}}\bigg]\bigg\}\,,
\nonumber \\ &&\!\!\!\!\!\!\!\!\!\!
q^{red}_{15} = -2q^{red}_{18} =
\frac{1}{48}Z^6_A\bigg[3\tilde{y}
           -2Z^2_A\bigg(5-12(1-Z^2_A)
           \frac{(1-\tilde{y})^2}{\tilde{y}}\bigg)\bigg]\,,
\nonumber \\ &&\!\!\!\!\!\!\!\!\!\!
q^{red}_{19} = \frac{1}{3}q^{red}_{18} =
-\frac{1}{192}Z^6_A(3\tilde{y}-2)\,,
\label{qred}
\end{eqnarray*}
where $\tilde{y} = 4\pi^2F_0^2/(Z^2_A N_c \overline{m}^2)=2.4$, and
$Z^2_A=0.62$ is the $\pi -A_1$ mixing factor.

    The current experimental status of the effective chiral lagrangian
at $O(p^4)$ has been discussed in some detail in \cite{dafne}.
    In table 6. we present the predictions of the NJL model for the structure 
coefficients $L_i$ and $H_2$ (without and with reduction of meson resonances)
in comparison with their phenomenological values.
    The predictions of the NJL model after reduction of meson resonances
turn out to be in a good agreement with phenomenology.

%===========================================================================
%                               References
%===========================================================================

%============================================================================
%\begin{center}
%\large {\bf Figure caption}
%\end{center}
%\section*{Figure caption}
%============================================================================
%
\hspace*{9mm}
\epsfysize=0.92\textheight
\epsfbox{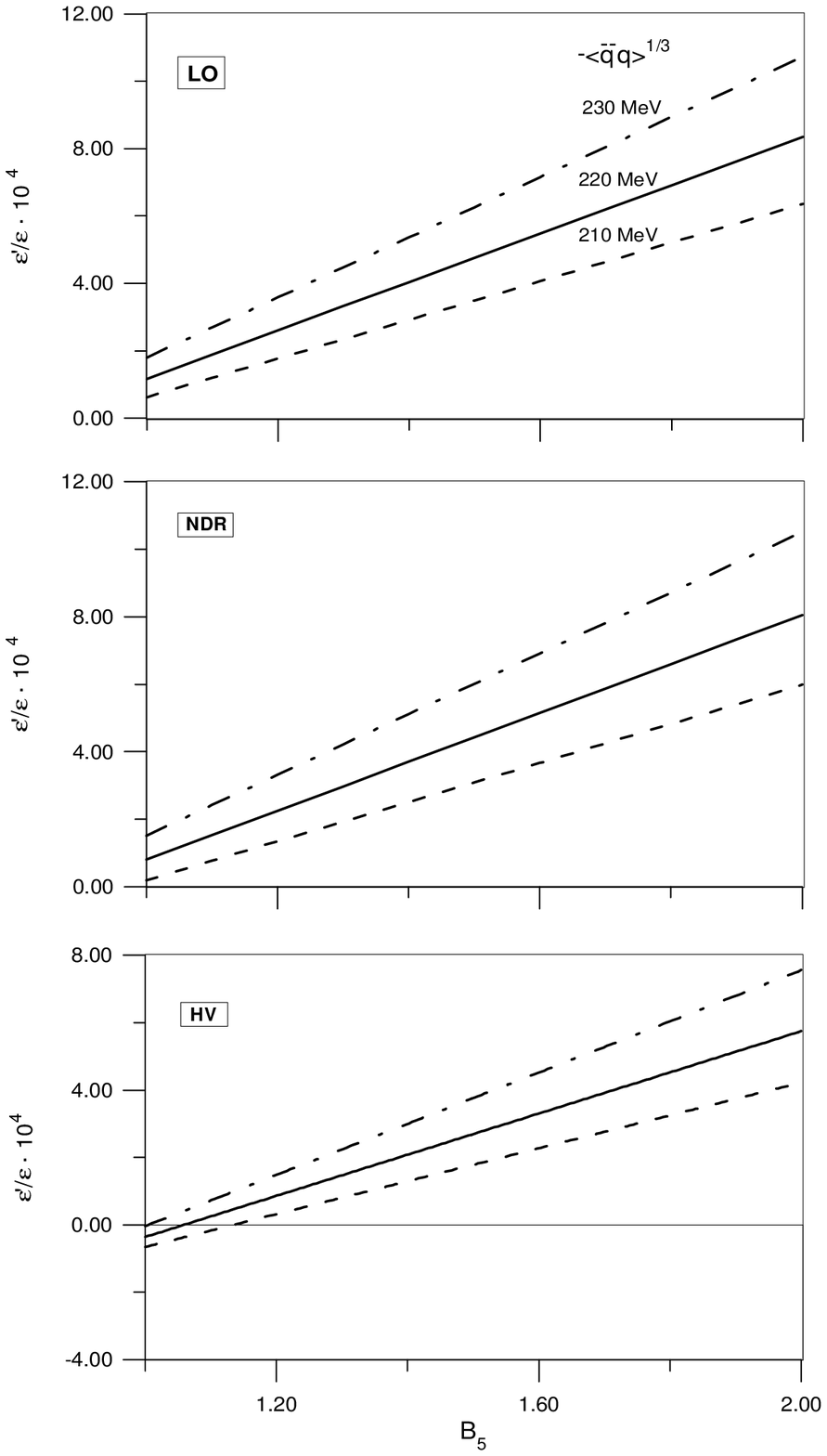}
\begin{center}
\large
{\bf Figure 1.}  $B_5$-dependence of $\varepsilon^{'}/\varepsilon$
                 for different values of the quark condensate,
                 $\Lambda ^{(4)}_{\overline{MS}} = 325$ MeV.
\end{center}
%
%
%============================================================================
%
\hspace*{9mm}
\epsfysize=0.92\textheight
\epsfbox{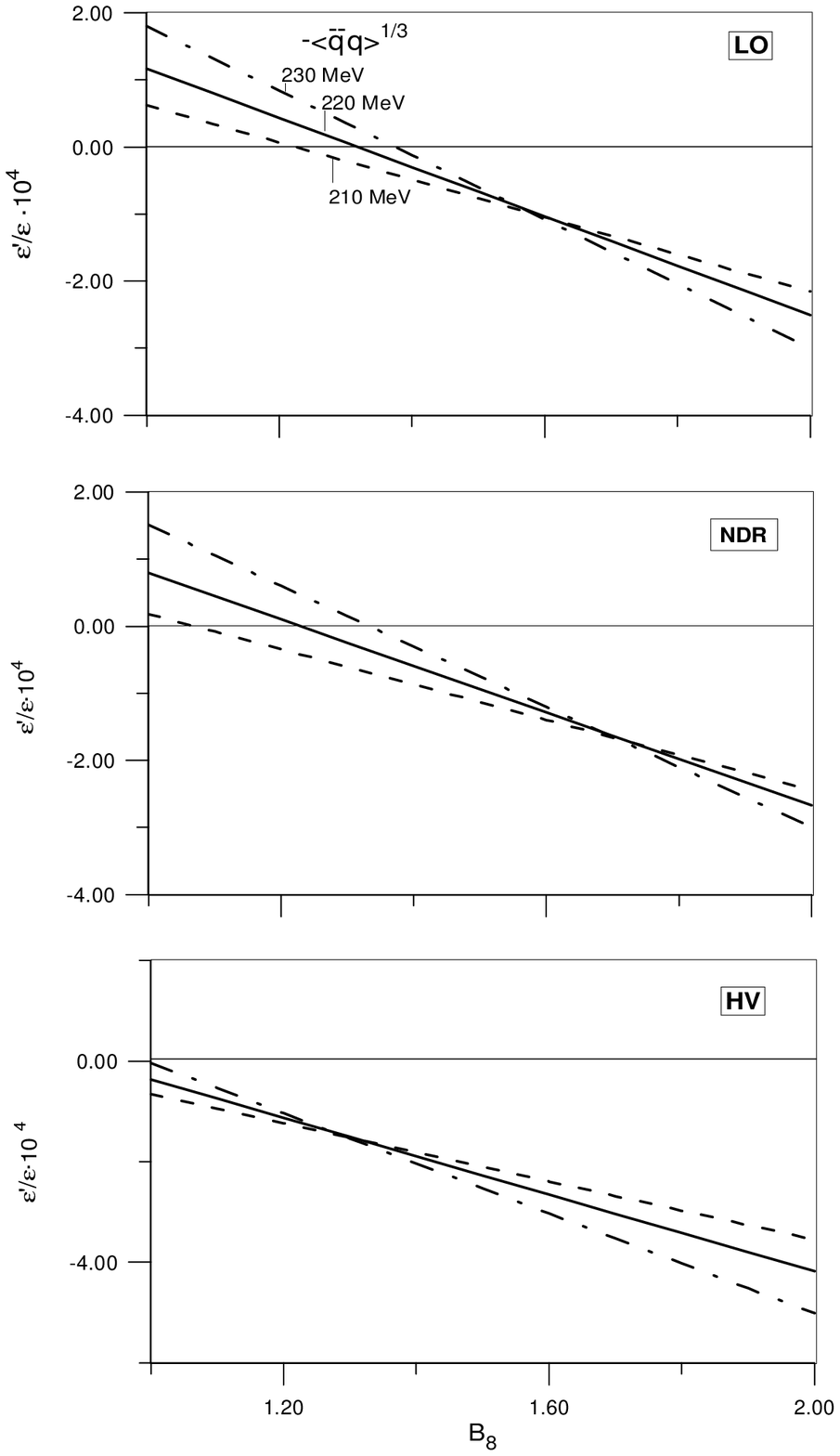}
\begin{center}
\large
{\bf Figure 2.}  $B_8$-dependence of $\varepsilon^{'}/\varepsilon$
                 for different values of the quark condensate,
                 $\Lambda ^{(4)}_{\overline{MS}} = 325$ MeV.
\end{center}
\newpage
%=============================================================================
%                                 Table 1
%=============================================================================
\begin{center}
\Large {\bf Table 1.}
\large Isotopic amplitudes of the $K\to2\pi$ decays\\
       with successive inclusion of higher-order corrections.\\
\end{center}
%------------------------------------------------------------------------
{\large~~~a) Without reduction of meson resonances:} \\[2mm]
%
%=============================================================================
%
\begin{tabular}{|l|l|*{8}{r}|} \hline
$~~~~~~~$&$~~~~~~~$&${\cal O}_1$ &${\cal O}_2$ &${\cal O}_3$ &${\cal O}_4$
                   &${\cal O}_5$ &${\cal O}_6$
                   &$\alpha {\cal O}_7$ &$\alpha {\cal O}_8$\\
\hline
$O(p^2)~$&$\mbox{Re} {\cal A}_0^{(i)}$
         &--1.000 & 1.000 & 1.000 & 0.000 &--6.833 & 0.000 & 0.016 & 1.015\\
$~~~~~~~$&$\mbox{Re} {\cal A}_2^{(i)}$
         & 0.000 & 0.000 & 0.000 & 1.000 & 0.019 & 0.000 &--0.016 & 0.454\\
\hline
\hline
$~~~~~~~$&$\mbox{Re} {\cal A}_0^{(i)}$
         &--1.184 & 1.173 & 1.090 & 0.021 &--7.836 & 0.003 & 0.016 & 1.206\\
$O(p^4)~$&$\mbox{Im} {\cal A}_0^{(i)}$
         &--0.482 & 0.482 & 0.482 & 0.000 &--3.229 & 0.000 & 0.008 & 0.534\\
$~~~~~~~$&$\mbox{Re} {\cal A}_2^{(i)}$
         &--0.005 & 0.016 & 0.035 & 0.867 &--0.077 &--0.003 &--0.015 & 0.424\\
$~~~~~~~$&$\mbox{Im} {\cal A}_2^{(i)}$
         & 0.000 & 0.000 & 0.000 &--0.213 &--0.004 & 0.000 & 0.003 &--0.064\\
\hline
\hline
$~~~~~~~$&$\mbox{Re} {\cal A}_0^{(i)}$
         &--1.014 & 1.003 & 0.888 & 0.022 &--6.413 &  0.003 & 0.012 & 1.042\\
$O(p^6)~$&$\mbox{Im} {\cal A}_0^{(i)}$
         &--0.707 & 0.709 & 0.682 & 0.000 &--4.582 &--0.001 & 0.011 & 0.740\\
$~~~~~~~$&$\mbox{Re} {\cal A}_2^{(i)}$
         &--0.004 & 0.016 & 0.035 & 0.812 &--0.078 &--0.003 &--0.015 & 0.396\\
$~~~~~~~$&$\mbox{Im} {\cal A}_2^{(i)}$
         & 0.000 & 0.001 & 0.001 &--0.114 & 0.001 & 0.000 & 0.002 &--0.054\\
\hline
\end{tabular}
\\[10mm]

%------------------------------------------------------------------------
{\large b) After reduction of resonances:} \\[2mm]
%
%=============================================================================
%
\begin{tabular}{|l|l|*{8}{r}|} \hline
$~~~~~~~$&$~~~~~~~$&${\cal O}_1$ &${\cal O}_2$ &${\cal O}_3$ &${\cal O}_4$
                   &${\cal O}_5$ &${\cal O}_6$
                   &$\alpha {\cal O}_7$ &$\alpha {\cal O}_8$\\
\hline
$O(p^2)~$&$\mbox{Re} {\cal A}_0^{(i)}$
         &--1.000 & 1.000 & 1.000 & 0.000 &--11.248& 0.000 & 0.016 & 1.325\\
$~~~~~~~$&$\mbox{Re} {\cal A}_2^{(i)}$
         & 0.000 & 0.000 & 0.000 & 1.000 & 0.017 & 0.000 &--0.016 & 0.574\\
\hline
\hline
$~~~~~~~$&$\mbox{Re} {\cal A}_0^{(i)}$
         &--1.195 & 1.187 & 1.101 & 0.021 &--13.524&  0.002 & 0.016 & 1.554\\
$O(p^4)~$&$\mbox{Im} {\cal A}_0^{(i)}$
         &--0.482 & 0.482 & 0.482 & 0.000 &--5.365 & 0.000 & 0.008 & 0.617\\
$~~~~~~~$&$\mbox{Re} {\cal A}_2^{(i)}$
         &--0.006 & 0.014 & 0.035 & 0.879 &--0.124 &-0.002 &--0.015 & 0.502\\
$~~~~~~~$&$\mbox{Im} {\cal A}_2^{(i)}$
         & 0.000 & 0.000 & 0.000 &--0.213 &--0.004 & 0.000 & 0.003 &--0.094\\
\hline
\hline
$~~~~~~~$&$\mbox{Re} {\cal A}_0^{(i)}$
         &--0.974 & 0.967 & 0.798 & 0.024 &--10.447 &  0.002 & 0.010 & 1.328\\
$O(p^6)~$&$\mbox{Im} {\cal A}_0^{(i)}$
         &--0.643 & 0.646 & 0.619 & 0.000 &--7.140 &--0.001 & 0.010 & 0.864\\
$~~~~~~~$&$\mbox{Re} {\cal A}_2^{(i)}$
         &-0.005 & 0.014 & 0.036 & 0.779 &--0.138 &--0.002 &--0.015 & 0.479\\
$~~~~~~~$&$\mbox{Im} {\cal A}_2^{(i)}$
         & 0.000 & 0.001 & 0.001 &--0.126 &--0.007 & 0.000 & 0.002 &--0.066\\
\hline
\end{tabular}
\newpage
%=============================================================================
%                                 Table 2
%=============================================================================
\begin{center}
\Large {\bf Table 2.}
\large QCD predictions for the parameters
       $\xi_i(\mu)=\xi^{(z)}_i(\mu)+\tau\xi^{(y)}_i(\mu)$ calculated
       with Wilson coefficients $c_i(\mu)=z_i(\mu)+\tau y_i(\mu)$
       at $\mu = 1$ GeV for $m_t = 170$ GeV \cite{buras3}.
\end{center}
\vspace{5mm}
\begin{tabular}{|l|r|r|r||r|r|r||r|r|r|}
\hline
\multicolumn{1}{|c|}{} &
\multicolumn{3}{|c||}{$\Lambda ^{(4)}_{\overline{MS}} = 215$ MeV \rule[-3mm]{0mm}{8mm}} &
\multicolumn{3}{|c||}{$\Lambda ^{(4)}_{\overline{MS}} = 325$ MeV} &
\multicolumn{3}{|c|}{$\Lambda ^{(4)}_{\overline{MS}} = 435$ MeV}\\ \hline
\multicolumn{1}{|c|}{$~~~~~~~~$} &
\multicolumn{1}{|c|}{~{\bf LO}~  }  &
\multicolumn{1}{|c|}{~{\bf NDR}} &
\multicolumn{1}{|c||}{~{\bf HV}~}  &
\multicolumn{1}{|c|}{~{\bf LO}~}  &
\multicolumn{1}{|c|}{~{\bf NDR}} &
\multicolumn{1}{|c||}{~{\bf HV}~}  &
\multicolumn{1}{|c|}{~{\bf LO}~}  &
\multicolumn{1}{|c|}{~{\bf NDR}} &
\multicolumn{1}{|c|}{~{\bf HV}~}\\ \hline
$\xi_1^{(z)}$       &--1.286 &--1.061 &--1.165 &--1.443 &--1.159 &
                     --1.325 &--1.624 &--1.270 &--1.562 \\
$\xi_2^{(z)}$       &  0.187 &  0.195 &  0.198 &  0.172 &  0.176 &
                       0.182 &  0.157 &  0.150 &  0.165 \\
$\xi_3^{(z)}$       &  0.129 &  0.143 &  0.137 &  0.122 &  0.137 &
                       0.130 &  0.115 &  0.131 &  0.121 \\
$\xi_4^{(z)}$       &  0.645 &  0.714 &  0.687 &  0.609 &  0.684 &
                       0.650 &  0.573 &  0.654 &  0.599 \\
$\xi_5^{(z)}$       &--0.008 &--0.020 &--0.008 &--0.012 &--0.032 &
                     --0.013 &--0.016 &--0.056 &--0.023 \\
$\xi_6^{(z)}$       &  0.000 &--0.003 &  0.000 &--0.001 &--0.007 &
                     --0.001 &--0.002 &--0.021 &--0.007 \\
$\xi_7^{(z)}/\alpha$&  0.002 &  0.003 &--0.001 &  0.004 &  0.008 &
                       0.001 &  0.006 &  0.015 &  0.032 \\
$\xi_8^{(z)}/\alpha$&  0.000 &  0.002 &  0.001 &  0.001 &  0.004 &
                       0.002 &  0.001 &  0.009 &  0.067 \\
\hline
\hline
$\xi_1^{(y)}$       &  0.044 &  0.038 &  0.048 &  0.054 &  0.048 &
                       0.053 &  0.065 &  0.060 &  0.069 \\
$\xi_2^{(y)}$       &--0.028 &--0.029 &--0.030 &--0.029 &--0.033 &
                     --0.030 &--0.030 &--0.033 &--0.030 \\
$\xi_3^{(y)}$       &--0.002 &--0.002 &  0.001 &--0.002 &--0.002 &
                     --0.002 &--0.002 &--0.002 &--0.002 \\
$\xi_4^{(y)}$       &--0.009 &--0.010 &  0.004 &--0.008 &--0.009 &
                     --0.009 &--0.008 &--0.009 &--0.008 \\
$\xi_5^{(y)}$       &--0.081 &--0.076 &--0.067 &--0.109 &--0.111 &
                     --0.092 &--0.143 &--0.173 &--0.132 \\
$\xi_6^{(y)}$       &--0.033 &--0.042 &--0.021 &--0.049 &--0.076 &
                     --0.033 &--0.071 &--0.139 &--0.051 \\
$\xi_7^{(y)}/\alpha$&  0.033 &  0.004 &  0.006 &  0.044 &  0.013 &
                       0.016 &  0.057 &  0.027 &  0.032 \\
$\xi_8^{(y)}/\alpha$&  0.031 &  0.028 &  0.031 &  0.043 &  0.041 &
                       0.045 &  0.058 &  0.061 &  0.067 \\
\hline
\end{tabular}
\newpage
\hoffset = -15mm
%=============================================================================
%                                 Table 3
%=============================================================================
\begin{center}
\Large {\bf Table 3.}
\large Predictions for the parameters of $K\to 2\pi$ decays\\
       in the semi-phenomenological approach ($B_5 = B_8 =1$).\\
       The ratio $\varepsilon^{'}/\varepsilon$ is given in units $10^{-4}$.
\end{center}
%=============================================================================
{\large ~~~a) Without reduction of meson resonances:} \\[4mm]
%
%=============================================================================
%
\small
\begin{tabular}{|c|c|c|c||c|c|c||c|c|c|}
\hline
\multicolumn{1}{|c|}{} &
\multicolumn{3}{|c||}{$\Lambda ^{(4)}_{\overline{MS}} = 215$ MeV} &
\multicolumn{3}{|c||}{$\Lambda ^{(4)}_{\overline{MS}} = 325$ MeV} &
\multicolumn{3}{|c|}{$\Lambda ^{(4)}_{\overline{MS}} = 435$ MeV\rule[-3mm]{0mm}{8mm}}
 \\ \hline
\multicolumn{1}{|c|}{$~~~~~~~~$} &
\multicolumn{1}{|c|}{$~~{\bf LO}~~$}  &
\multicolumn{1}{|c|}{$~~{\bf NDR}~$} &
\multicolumn{1}{|c||}{$~~{\bf HV}~~$}  &
\multicolumn{1}{|c|}{$~~{\bf LO}~~$}  &
\multicolumn{1}{|c|}{$~~{\bf NDR}~$} &
\multicolumn{1}{|c||}{$~~{\bf HV}~~$}  &
\multicolumn{1}{|c|}{$~~{\bf LO}~~$}  &
\multicolumn{1}{|c|}{$~~{\bf NDR}~$} &
\multicolumn{1}{|c|}{$~~{\bf HV}~~$}\\ \hline
$B_{1}$
& 5.52 & 6.28 & 5.90 & 5.07 & 5.91 & 5.39 & 4.63 & 5.52 & 4.73 \\
$B_{4}$
& 0.65 & 0.59 & 0.62 & 0.69 & 0.61 & 0.65 & 0.74 & 0.64 & 0.70 \\
$P_0$
& 0.95 & 0.45 &--0.09 & 2.09 & 1.63 & 1.12 & 3.59 & 4.21 & 2.96 \\
$P_2$
& 1.71 & 1.46 & 2.64 & 2.76 & 2.65 & 2.68 & 4.13 & 4.65 & 4.28 \\
$(\varepsilon^{'}/\varepsilon )_{min}$
&--1.3 &--1.7 &--4.7 &--1.2 &--1.7 &--2.7 &--0.9 &--0.8 &--2.3 \\
$(\varepsilon^{'}/\varepsilon )_{max}$
&--0.7 &--0.9 &--2.4 &--0.6 &--0.9 &--1.3 &--0.5 &--0.4 &--1.1 \\
\hline
\end{tabular}
\\[10mm]

%------------------------------------------------------------------------
{\large b) After reduction of resonances:} \\[4mm]
%
%=============================================================================
\small
\begin{tabular}{|c|c|c|c||c|c|c||c|c|c|}
\hline
\multicolumn{1}{|c|}{} &
\multicolumn{3}{|c||}{$\Lambda ^{(4)}_{\overline{MS}} = 215$ MeV} &
\multicolumn{3}{|c||}{$\Lambda ^{(4)}_{\overline{MS}} = 325$ MeV} &
\multicolumn{3}{|c|}{$\Lambda ^{(4)}_{\overline{MS}} = 435$ MeV\rule[-3mm]{0mm}{8mm}}
 \\ \hline
\multicolumn{1}{|c|}{$~~~~~~~~$} &
\multicolumn{1}{|c|}{$~~{\bf LO}~~$}  &
\multicolumn{1}{|c|}{$~~{\bf NDR}~$} &
\multicolumn{1}{|c||}{$~~{\bf HV}~~$}  &
\multicolumn{1}{|c|}{$~~{\bf LO}~~$}  &
\multicolumn{1}{|c|}{$~~{\bf NDR}~$} &
\multicolumn{1}{|c||}{$~~{\bf HV}~~$}  &
\multicolumn{1}{|c|}{$~~{\bf LO}~~$}  &
\multicolumn{1}{|c|}{$~~{\bf NDR}~$} &
\multicolumn{1}{|c|}{$~~{\bf HV}~~$}\\ \hline
$B_{1}$
& 5.85 & 6.62 & 6.25 & 5.36 & 6.20 & 5.69 & 4.88 & 5.72 & 4.97 \\
$B_{4}$
& 0.68 & 0.61 & 0.64 & 0.72 & 0.64 & 0.67 & 0.76 & 0.66 & 0.73 \\
$P_0$
& 3.21 & 2.56 & 1.79 & 5.12 & 4.69 & 3.72 & 7.59 & 8.94 & 6.67 \\
$P_2$
& 2.77 & 2.43 & 3.54 & 4.21 & 4.07 & 4.00 & 6.06 & 6.85 & 6.05 \\
$(\varepsilon^{'}/\varepsilon )_{min}$
& 0.4 & 0.1 &--3.0 & 0.8 & 0.5 &--0.5 & 1.3 & 1.8 & 0.5 \\
$(\varepsilon^{'}/\varepsilon )_{max}$
& 0.8 & 0.2 &--1.5 & 1.6 & 1.1 &--0.2 & 2.6 & 3.6 & 1.1 \\
\hline
\end{tabular}
\newpage

%=============================================================================
%                                 Table 4
%=============================================================================
\begin{center}
\Large {\bf Table 4.}
\large Dependence of the predictions for the ratio
       $\varepsilon^{'}/\varepsilon$\\
       (in units $10^{-4}$) on the factor $B_5$ ($B_8 =1$).\\
\end{center}
%=============================================================================
\small
\begin{tabular}{|c|c|c|c||c|c|c||c|c|c|}
\hline
\multicolumn{1}{|c|}{} &
\multicolumn{3}{|c||}{$\Lambda ^{(4)}_{\overline{MS}} = 215$ MeV} &
\multicolumn{3}{|c||}{$\Lambda ^{(4)}_{\overline{MS}} = 325$ MeV} &
\multicolumn{3}{|c|}{$\Lambda ^{(4)}_{\overline{MS}} = 435$ MeV\rule[-3mm]{0mm}{8mm}}
\\ \hline
\multicolumn{1}{|c|}{$~~B_5~~$} &
\multicolumn{1}{|c|}{$~~{\bf LO}~~$}  &
\multicolumn{1}{|c|}{$~~{\bf NDR}~$} &
\multicolumn{1}{|c||}{$~~{\bf HV}~~$}  &
\multicolumn{1}{|c|}{$~~{\bf LO}~~$}  &
\multicolumn{1}{|c|}{$~~{\bf NDR}~$} &
\multicolumn{1}{|c||}{$~~{\bf HV}~~$}  &
\multicolumn{1}{|c|}{$~~{\bf LO}~~$}  &
\multicolumn{1}{|c|}{$~~{\bf NDR}~$} &
\multicolumn{1}{|c|}{$~~{\bf HV}~~$}\\ \hline
$ 1.0 $&  0.56 &  0.16 &--2.26 &  1.17 &  0.80 &--0.36 &  1.97 &  2.70 &  0.80 \\
$ 1.1 $&  1.10 &  0.67 &--1.81 &  1.89 &  1.52 &  0.25 &  2.92 &  3.82 &  1.70 \\
$ 1.2 $&  1.64 &  1.17 &--1.37 &  2.60 &  2.25 &  0.86 &  3.86 &  4.94 &  2.59 \\
$ 1.3 $&  2.17 &  1.67 &--0.92 &  3.32 &  2.97 &  1.47 &  4.81 &  6.06 &  3.48 \\
$ 1.4 $&  2.71 &  2.17 &--0.48 &  4.04 &  3.70 &  2.08 &  5.75 &  7.18 &  4.38 \\
$ 1.5 $&  3.25 &  2.67 &--0.03 &  4.76 &  4.42 &  2.69 &  6.69 &  8.30 &  5.27 \\
$ 1.6 $&  3.78 &  3.17 &  0.42 &  5.47 &  5.15 &  3.30 &  7.64 &  9.42 &  6.16 \\
$ 1.7 $&  4.32 &  3.68 &  0.86 &  6.19 &  5.87 &  3.92 &  8.58 & 10.53 &  7.06 \\
$ 1.8 $&  4.86 &  4.18 &  1.31 &  6.91 &  6.60 &  4.53 &  9.52 & 11.65 &  7.95 \\
$ 1.9 $&  5.39 &  4.68 &  1.75 &  7.63 &  7.32 &  5.14 & 10.47 & 12.76 &  8.84 \\
$ 2.0 $&  5.93 &  5.18 &  2.20 &  8.34 &  8.05 &  5.75 & 11.41 & 13.88 &  9.73 \\
\hline
\end{tabular}
%\newpage

\vspace{10mm}
%=============================================================================
%                                 Table 5
%=============================================================================
\begin{center}
\Large {\bf Table 5.}
\large Dependence of the predictions for the ratio
       $\varepsilon^{'}/\varepsilon$\\
       (in units $10^{-4}$) on the factor $B_8$ ($B_5 =1$).\\
\end{center}
%=============================================================================
\small
\begin{tabular}{|c|c|c|c||c|c|c||c|c|c|}
\hline
\multicolumn{1}{|c|}{} &
\multicolumn{3}{|c||}{$\Lambda ^{(4)}_{\overline{MS}} = 215$ MeV} &
\multicolumn{3}{|c||}{$\Lambda ^{(4)}_{\overline{MS}} = 325$ MeV} &
\multicolumn{3}{|c|}{$\Lambda ^{(4)}_{\overline{MS}} = 435$ MeV\rule[-3mm]{0mm}{8mm}}
\\ \hline
\multicolumn{1}{|c|}{$~~B_8~~$} &
\multicolumn{1}{|c|}{$~~{\bf LO}~~$}  &
\multicolumn{1}{|c|}{$~~{\bf NDR}~$} &
\multicolumn{1}{|c||}{$~~{\bf HV}~~$}  &
\multicolumn{1}{|c|}{$~~{\bf LO}~~$}  &
\multicolumn{1}{|c|}{$~~{\bf NDR}~$} &
\multicolumn{1}{|c||}{$~~{\bf HV}~~$}  &
\multicolumn{1}{|c|}{$~~{\bf LO}~~$}  &
\multicolumn{1}{|c|}{$~~{\bf NDR}~$} &
\multicolumn{1}{|c|}{$~~{\bf HV}~~$}\\ \hline
$ 1.0 $&  0.56 &  0.16 &--2.26 &  1.17 &  0.79 &--0.36 &  1.97 &  2.70 &  0.80 \\
$ 1.1 $&  0.30 &--0.07 &--2.52 &  0.80 &  0.45 &--0.75 &  1.48 &  2.19 &  0.32 \\
$ 1.2 $&  0.03 &--0.31 &--2.78 &  0.43 &  0.10 &--1.13 &  0.98 &  1.67 &--0.15 \\
$ 1.3 $&--0.23 &--0.55 &--3.05 &  0.07 &--0.25 &--1.51 &  0.48 &  1.15 &--0.61 \\
$ 1.4 $&--0.50 &--0.78 &--3.31 &--0.30 &--0.60 &--1.89 &--0.02 &  0.64 &--1.07 \\
$ 1.5 $&--0.77 &--1.02 &--3.57 &--0.67 &--0.94 &--2.28 &--0.52 &  0.12 &--1.52 \\
$ 1.6 $&--1.03 &--1.25 &--3.84 &--1.04 &--1.29 &--2.66 &--1.02 &--0.39 &--1.96 \\
$ 1.7 $&--1.30 &--1.49 &--4.10 &--1.40 &--1.63 &--3.04 &--1.52 &--0.90 &--2.40 \\
$ 1.8 $&--1.56 &--1.73 &--4.36 &--1.77 &--1.98 &--3.42 &--2.01 &--1.42 &--2.82 \\
$ 1.9 $&--1.83 &--1.96 &--4.62 &--2.14 &--2.32 &--3.80 &--2.51 &--1.92 &--3.25 \\
$ 2.0 $&--2.09 &--2.20 &--4.89 &--2.51 &--2.67 &--4.18 &--3.01 &--2.43 &--3.66 \\
\hline
\end{tabular}
\newpage
%=============================================================================
%                            Table 6
%=============================================================================
\begin{center}
\Large {\bf Table 6.}
\large Theoretical and phenomenological values of the structure coefficients
       $L_i$ and $H_2$
\footnote{The parameter $H_2$ has not been determined phenomenologically
          until now.}
 (in units $10^{-3}$).
\\
\end{center}
%=============================================================================
\begin{center}
\begin{tabular}{|c|c|c|c|}
\hline
Structure   & Without reduction & After reduction &
Phenomenology \cite{{dafne}}\\
coefficients&   of resonances   & of resonances   & \\
\hline
$L_1$ & 0.79 & 0.85 & $ 0.4 \pm 0.3 $\\
$L_2$ & 1.58 & 1.70 & $ 1.4 \pm 0.3 $\\
$L_3$ &-3.17 &-4.30 & $-3.5 \pm 1.1 $\\
$L_4$ &  0   &  0   & $-0.3 \pm 0.5 $\\
$L_5$ & 0.98 & 1.64 & $ 1.4 \pm 0.5 $\\
$L_8$ & 0.36 & 1.12 & $ 0.9 \pm 0.3 $\\
\hline
$H_2$ & 1.01 &-0.67 & $             $\\
\hline 
\end{tabular}
\end{center}
%-----------------------------------------------------------------------------
\end{document}